\documentclass[useAMS,usegraphicx,usenatbib,usedcolumn]{mn2e}
\usepackage{subfigure}

\title[Radio emission from dark matter annihilation in the LMC]{Radio emission from dark matter annihilation in the Large Magellanic Cloud}
\author[Beatriz B. Siffert et al.]{Beatriz B. Siffert$^{1}$\thanks{E-mail: siffert@na.infn.it}, Angelo Limone$^{1}$, Enrico Borriello$^{1,2}$, Giuseppe Longo$^{1,3,4}$ \newauthor and Gennaro Miele$^{1,3}$\\
$^{1}$Dipartimento di Scienze Fisiche, Universit\`a di Napoli Federico II, Via Cinthia, 9, I-80126, Naples, Italy\\
$^{2}$INFN - Sezione di Napoli, Via Cinthia, 9, I-80126, Naples, Italy\\
$^{3}$INAF-OACN, Via Moiariello, 16, I-80128, Naples, Italy\\
$^{4}$Department of Astronomy, California Institute of Technology, 1200 East California Blvd, Pasadena, CA 91125, USA}
\begin{document}


\pagerange{\pageref{firstpage}--\pageref{lastpage}} \pubyear{0000}

\maketitle

\label{firstpage}

\begin{abstract}
The Large Magellanic Cloud, at only 50 kpc away from us and known to be dark matter dominated, is clearly an interesting place where to search for dark matter annihilation signals. In this paper, we estimate the synchrotron emission due to WIMP annihilation in the halo of the LMC at two radio frequencies, 1.4 and 4.8 GHz, and compare it to the observed emission, in order to impose constraints in the WIMP mass vs. annihilation cross section plane. We use available Faraday rotation data from background sources to estimate the magnitude of the magnetic field in different regions of the LMC's disc, where we calculate the radio signal due to dark matter annihilation. We account for the $e^+e^-$ energy losses due to synchrotron, Inverse Compton Scattering and bremsstrahlung, using the observed hydrogen and dust temperature distribution on the LMC to estimate their efficiency. The extensive use of observations, allied with conservative choices adopted in all the steps of the calculation, allow us to obtain very realistic constraints. 
  
\end{abstract}


\section{Introduction}
Evidence for the existence of dark matter (DM) has been observed in various astrophysical systems, ranging from satellite dwarf galaxies to massive galactic clusters and cosmology. Possible cold dark matter candidates arise from theoretical models conceived to extend the Standard Model of elementary particles and Weakly Interacting Massive Particles (WIMP) are the current main paradigm (see \citet{RevDM} for a recent review on dark matter candidates).  

The fact that direct observation of dark matter particles has not yet been possible justifies all the efforts devoted to their indirect detection, i.e., the observation of anomalous components in the cosmic rays' spectrum that can be attributed to dark matter annihilation. See \citet{gamma_radio,gammaI,Xray,radioI,radioII,neutrino,multiI,multiII} for examples of such searches in the $\gamma$-ray, X-ray, radio and neutrino spectra.

In fact, dark matter self-annihilation is expected to produce several Standard Model particles, among which electrons and positrons that, by interacting with the galactic magnetic field, emit synchrotron radiation in the radio band. We can estimate the intensity of the emission coming from a given direction in the sky by adopting models or using measurements to describe the following quantities: the dark matter density profile, which can be modelled by using the results of halo formation numerical simulations; the magnetic field, estimated through techniques such as the analysis of polarisation of radio and optical emission and rotation measures; and the interstellar radiation field and hydrogen distribution, which are needed to account for the electron/positron energy losses. The comparison of this emission with the observed radio emission then allows to impose constraints on the values of the dark matter's mass, $m_{\chi}$, and its thermally averaged annihilation cross section, $\langle \sigma_A v \rangle$. 

In order for this method to provide reliable constraints, the values of all the involved parameters need to be known very well, which often does not happen. Furthermore, ideally, the comparison between the theoretical result and the observed emission should be made after all the known astrophysical foreground has been subtracted from the observations. In practice, however, such subtraction is subject to several uncertainties, since often we cannot accurately model all the components contributing to the foreground. 

In addition, the method presents an intrinsic shortcoming: we can only calculate the contributions from DM annihilation integrated along the line of sight, so all the DM signal may be `averaged' away by low emitting regions that happen to fall along the same direction in the sky.  

In the present paper, we present constraints obtained by applying this method to the Large Magellanic Cloud (LMC) and comparing the results with radio observations at 1.4 and 4.8 GHz. Unless not otherwise stated, we will follow the formalism described in \citet*{PaperMW} in our calculations. We analyse two possible WIMP annihilation channels, a hadronic and a leptonic one, the latter having been recently proposed as a possible cause of the anomalous cosmic ray electron/positron spectrum measured by Pamela \citep{PAMELA} and Fermi--LAT \citep{b1}.

By extensive use of available observations of the LMC in different frequencies, we are able to obtain realistic information to describe all the inputs necessary to estimate the DM annihilation signal. In this way, we escape from the `too many hypotheses' problem that has recurrently appeared in the existing literature on the subject. When we do have to make a hypothesis, we follow the most conservative path, making sure that our choice will not overestimate the signal.

The LMC, at a distance of only $\sim 50$ kpc from us \citep{distLMC}, is probably one of the best studied galaxies in almost all frequency bands. Rotation curve data imply that the LMC total mass inside a 9 kpc radius is $\sim 1.3 \times 10^{10}$ M$_{\odot}$, while the sum of the stellar and neutral gas masses is estimated to be $\sim 3.2 \times 10^{9}$ M$_{\odot}$ \citep{VanDerMarelProceedings}. Therefore, the LMC must be dark matter dominated.

The fact that the LMC is nearly face-on (its disc forms an angle of $\sim 35^{\circ}$ with the plane of the sky \citep{VanDerMarelInclination}), minimises the line-of-sight integration problem mentioned above, since, in this case, the integration spans only over the thickness of the galaxy. A nearly face-on view also allows us to clearly distinguish between high emitting and low emitting regions within the disc, i.e. between regions where different astrophysical processes are taking place and different astrophysical foregrounds exist. 

In \citet{PaperM33}, some of the authors of the present paper looked for DM annihilation from low radio emitting regions (called `radio cavities') within the nearly face-on Local Group member Messier 33 (inclination  $56^{\circ}$). Following this reference, a good candidate radio cavity should present low radio emission relative to the rest of the disc and should not be very far from the centre of the galaxy. While the first criterion guarantees that the astrophysical foreground in that region is relatively small, the second one is required so that the DM density is non negligible and that the possibility of a high magnetic field is not ruled out. Although this study produced constraints comparable to those obtained for the Milky Way (see for example \citet{PaperMW}), the lack of knowledge of the values of the astrophysical environment governing the diffusion of the electrons and positrons, in particular the magnetic field, introduces several uncertainties on the results.

In the current analysis, in contrast, we use recent accurate rotation measures (RMs) from point sources behind the LMC obtained in \citet{MagField}, through which we can directly estimate the magnitude of the magnetic field in several regions for the disc of the LMC. We focus our analysis on 23 regions on the LMC's disc with galactocentric distances $\la$ 8 kpc for which a RM has been measured. Although these regions are not chosen to be radio cavities, the fact that a background radio source has been seen there, guarantees that they present low radio emission in comparison to other parts of the galaxy (intensity at 1.4 and 4.8 GHz $\la$ 2 mJy/beam). For each of these regions we calculate the expected DM emission at 1.4 and 4.8 GHz and compare the results with the observed values without performing any foreground subtraction (which is a conservative choice, since removing part of the observed emission would always strengthen the constraints).

Previous searches for dark matter annihilation signals from the LMC have analysed both the $\gamma$-ray \citep{Gondolo, Angela} and radio \citep*{Angela} bands. In particular, in the latter the flux density in several radio frequencies due to annihilation into the hadronic channel integrated over the entire galaxy is calculated. However, unlike what is done here, in \citet{Angela} a simple constant magnetic field inside the entire volume of the galaxy is assumed. Furthermore, electron/positron energy losses due to Inverse Compton Scattering and bremsstrahlung are neglected in \citet{Angela} although, as we show in section~\ref{EnergyLosses}, these processes can be very important in several regions of the LMC. When appropriate, we will comment on the effect of these assumptions and compare our results with theirs.

The paper is organised as follows. In section~\ref{RadioData} we present the radio data used in our analysis and the 23 regions studied. Section~\ref{profiles} is dedicated to the distribution of dark matter in the halo of the LMC and presents the results of fits to different dark matter profiles. In section~\ref{EnergyLosses} we discuss the energy loss processes important to our calculations. In section~\ref{ImposingConstraints} we present our results and discuss them and compare with previous ones in section~\ref{Discussion}.

\section{Radio data and selection of regions} \label{RadioData}
\begin{figure*}
\includegraphics[scale=0.4]{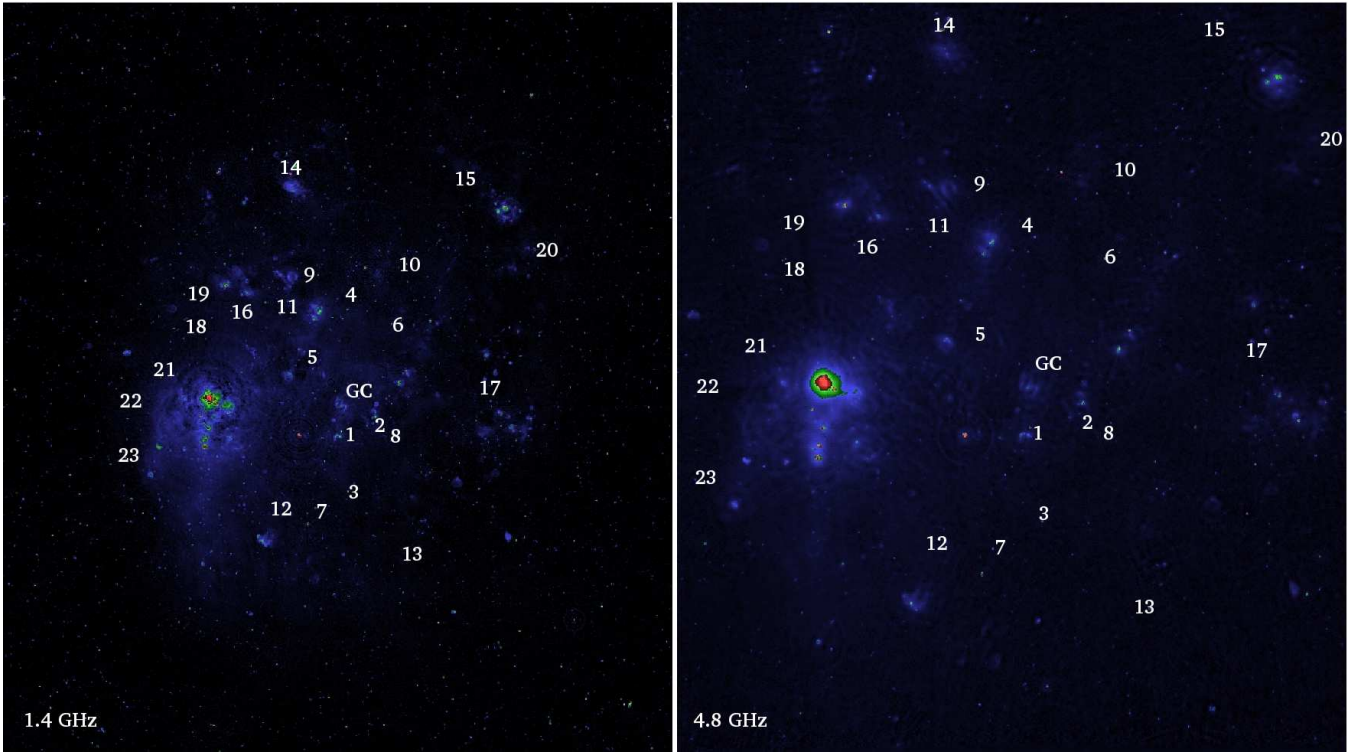}
\caption{Radio maps in logarithmic scale of the LMC at 1.4 GHz \protect\citep{RadioData1dot4} and 4.8 GHz \protect\citep{RadioData4dot8}. The numbers indicate the positions of the 23 selected regions where the Faraday rotation of background sources is measured. The label `GC' indicates the assumed LMC's kinematic centre.} \label{RadioMaps}
\end{figure*}

We used radio observations of the LMC at two different frequencies, 1.4 and 4.8 GHz. At 1.4 GHz, we used the mosaic image obtained with the Australia Telescope Compact Array (ATCA) and the Parkes Telescope \citep{RadioData1dot4}, shown in the left panel of Fig.~\ref{RadioMaps}. The HPBW for this image is 40 arcsec, which corresponds to $\simeq 10$ pc at the LMC's position, and its sensitivity is estimated to be between 0.25 and 0.35 mJy/beam. At 4.8 GHz, we used the observations made with the ATCA described in \citet{RadioData4dot8} and shown in the right panel of Fig.~\ref{RadioMaps}. The HPBW is 33 arcsec and the estimated sensitivity is 0.28 mJy/beam. 

We estimated the radio emission due to dark matter annihilation at these two frequencies coming from 23 circular regions with radius 180 arcsec within the LMC's disc. The selection of regions was made according to the results presented in \citet{MagField}, where the Faraday rotation for 291 polarised sources behind the LMC was measured, 100 of which happen to lay directly behind the LMC. After subtracting the effects due to the foreground Faraday rotation in the Milky Way, a map of rotation measures is obtained (fig. 1 in \citet{MagField}), where circles of various sizes represent the magnitude of the RM at each position. We chose the positions of 23 circles to apply our method, selecting them according to their diameter, which is proportional to the RM at that position (large RM implies large magnetic field and therefore large synchrotron emission), and distance from the centre of the LMC (as we move away from the centre the density of DM decreases). In our calculations we used the LMC's kinematic centre, with coordinates RA$\,=\,05\,\rm{h}\,17\,\rm{m}\,36\,\rm{s}$ and DEC$\,=\,-69^{\circ}$02' \citep{H1RotationCurve}.

\begin{table*}
\caption{The 23 regions studied. Intensity is measured in mJy/beam.}  \label{TableRCs}
\begin{tabular}{@{}lcccccccccc@{}}
\hline
 & RA$\,$(J2000) & DEC$\,$(J2000) & Name & d$\,$(kpc) & B$_{{\rm reg}}\,$($\mu$G) & B$_{{\rm tot}}\,$($\mu$G) & U$_{{\rm rad}}^{{\rm disc}}\,$(eV$\,$cm$^{-3}$) & n$_{\rm H}\,$(cm$^{-3}$) & I$_{1.4\,{\rm GHz}}$ & I$_{4.8\,{\rm GHz}}$\\
\hline
1 & 05:17:43.541 & -69:35:17.07 & MDM3 & 0.56 & 2.45 & 9.14  & 1.508 & 0.07 & 1.83 & 1.58\\
2 & 05:13:14.664 & -69:31:17.19 & - & 1.21 &  2.76 & 10.33  & 1.346 & 0.14 & 1.67 & 1.27\\
3 & 05:17:22.087 & -70:20:41.31 & MDM1  & 1.33 & 1.59 &  5.96  & 1.098 & 0.07 & 1.56 & 1.01\\
4 & 05:18:51.899 & -67:45:54.21 & MDM8  & 1.36 &  3.61 & 13.51 & 1.030 & 0.07 & 0.94 & 0.67\\
5 & 05:23:46.213 & -68:45:33.88 & MDM20 & 1.51 & 1.70 & 6.36 & 1.207 & 0.14 & 2.18 & 1.47\\
6 & 05:10:45.796 & -68:05:11.09 & - & 1.71 &  2.23 & 8.34  & 1.135 & 0.07 & 1.40 & 0.69\\
7 & 05:21:50.892 & -70:35:46.96 & MDM18 & 1.71 & 2.45 & 9.14  & 0.439 & 0.07 & 0.84 & 0.43\\
8 & 05:11:01.258 & -69:34:40.98 & - & 1.72 &  2.66 & 9.93 & 0.857 & 0.14 & 1.06 & 0.60\\
9 & 05:22:57.600 & -67:29:16.80 & MDM17 & 2.17 &  5.32 & 19.86 & 0.921 & 0.07 & 0.80 & 0.48\\
10 & 05:10:21.596 & -67:17:12.72 & - & 2.21 &   3.08 & 11.52 & 0.997 & 0.54 & 1.52 & 0.58\\
11 & 05:27:09.571 & -67:49:06.69 & MDM30 & 2.73 & 2.34 & 8.74  & 1.266 & 1.93 & 2.09 & 1.69\\
12 & 05:28:40.638 & -70:35:34.72 & MDM32 & 2.78 &  2.23 & 8.34 & 1.233 & 0.14 & 1.50 & 1.23\\
13 & 05:06:19.726 & -71:08:26.64 & - & 3.67 & 4.04 & 15.10  & 0.637 & 0.07 & 0.63 & 0.28*\\
14 & 05:26:09.659 & -66:00:02.93 & - & 3.95 &  4.25 & 15.89 & 0.569 & 0.42 & 1.63 & 1.73\\
15 & 05:02:44.925 & -66:00:28.26 & - & 4.19 & 4.57 & 17.08  & 0.643 & 0 & 0.68 & 0.36\\
16 & 05:34:13.644 & -67:55:04.42 & MDM58 & 4.23 & 2.23 & 8.34   & 1.106 & 0.14 & 2.11 & 1.38\\
17 & 04:56:57.832 & -68:50:35.85 & - & 4.79 & 1.70 & 6.36 & 0.985 & 0.24 & 1.00 & 0.78\\
18 & 05:41:11.743 & -68:03:33.24 & MDM81 & 5.77 &  3.72 & 13.90  & 0.601 & 0.14 & 0.78 & 0.38\\
19 & 05:40:54.847 & -67:41:04.43 & MDM82 & 5.85 &  3.72 & 13.90  & 0.694 & 0.14 & 1.12 & 0.53\\
20 & 04:51:36.413 & -66:51:26.34 & - & 6.07 &  2.34 & 8.74 & 0.900 & 0 & 0.64 & 0.28*\\
21 & 05:45:23.757 & -68:42:57.08 & MDM95 & 6.56 &  2.76 & 10.33  & 0.878 & 0.14 & 1.35 & 0.53\\
22 & 05:50:19.691 & -69:00:39.60 & - & 7.64 &  2.45 & 9.14  & 0.494 & 0.14 & 1.00 & 0.62\\
23 & 05:51:58.795 & -69:45:06.22 & MDM113  & 7.92 &  2.45  & 9.14  & 0.461 & 0.14 & 0.46 & 0.28*\\
\hline
\end{tabular}
\end{table*}

Since the data published in \citet{MagField} are not yet of public domain, we had to obtain the positions and the RM of each of the 23 regions directly from the map in their fig. 1, using the fact that the maximum RM measured is $+247 \pm 13$ rad m$^{-2}$. To help us determining the coordinates of each region, we used the list of sources published in \citet{BackgroundSource}, which contains 113 compact radio sources detected with the ATCA at 1.4 GHz in and behind the LMC, in a region comprehended between $05\,\rm{h}\,17\,\rm{m} \la \rm{RA} \la 05\,\rm{h}\,51\,\rm{m}$ and $-71^{\circ} \la\rm{DEC} \la -67^{\circ}$. It is estimated that among these 113 sources, 15 are in the LMC, so most of them are background objects.  

Once RMs have been estimated for each region, the corresponding magnetic field was obtained using the definition:
\begin{equation}
{\rm RM} = k \int n_e \,B_{{\rm reg}\parallel}\,ds,
\end{equation}
where $k=0.81$ rad m$^{-2}$pc$^{-1}$cm$^3$ $\mu\,$G$^{-1}$, $n_e$ is the electron number density, $B_{{\rm reg}\parallel}$ is the regular component of the magnetic field parallel to line-of-sight and $s$ is the distance along the line-of-sight. 

Assuming that the regular component of the magnetic field, $B_{{\rm reg}}$, lies solely on the plane of the disc, we conclude that $B_{{\rm reg}\parallel}$ is produced solely by the galaxy's inclination. The mean value of $B_{{\rm reg}}$, weighted according to the electron density $n_e$ along the line-of-sight is then given by
\begin{equation}
B_{{\rm reg}}= \frac{{\rm RM}}{k\,\,{\rm \mathcal{DM}}\,{\rm sin}i},
\end{equation}
where $\mathcal{DM}$ is the dispersion measure, defined as $\int n_e\, ds$ and $i\sim35^{\circ}$ is the LMC's disc inclination. In our calculations we used the mean total dispersion measure calculated from five radio pulsars in the LMC obtained in \citet{DispersionMeasure}, $\mathcal{DM}=100$ cm$^{-3}$ pc, following what was done in \citet{MagField}.

In \citet{MagField}, the random component of the LMC's magnetic field, $B_{{\rm ran}}$, was estimated to be $\sim3.6\,B_{{\rm reg}}$, so that $B_{{\rm tot}}=\sqrt{B_{{\rm reg}}^2+B_{{\rm ran}}^2} \simeq 3.7 B_{\rm reg}$. As mentioned in \citet{MagField}, in regions where the magnetic field and $n_e$ are correlated, the above calculation overestimates $B_{{\rm reg}}$ by a factor 2 and underestimates $B_{{\rm ran}}$ by the same factor.

The 23 selected regions are shown in Fig.~\ref{RadioMaps}, superimposed to the radio maps. They are numbered from the smallest to the largest galactocentric distance. In Table~\ref{TableRCs} we list for each region: its equatorial coordinates; corresponding background source name as in \citet{BackgroundSource}, if any; distance to the centre of the LMC along the plane of the disc, assuming a distance to the LMC of 50 kpc; regular and total magnetic fields estimated from the RM; interstellar radiation field on the disc (see section~\ref{EnergyLosses}); the atomic hydrogen number density (see section~\ref{EnergyLosses}); and intensity measured at 1.4 and 4.8 GHz. The values of intensity at 4.8 GHz marked with an asterisk were lower than the corresponding sensitivity and were set to the sensitivity value, 0.28 mJy/beam.

To model the behaviour of the magnetic field along the z-axis (perpendicular to the disc plane) we use an exponential decay of the following form:
\begin{equation}
B(z)=B_{{\rm disc}}\,{\rm e}^{-|z|/z_0}, \label{BExponential}
\end{equation}
where $z_0$ is the scale height of the magnetic field, which can be up to 4 times larger than the scale height of the thick synchrotron disc (in the case of equipartition between cosmic rays and magnetic fields and a synchrotron spectral index $\simeq$1) \citep{BScaleHeightBeck}. The thick disc has typically a scale height of $\sim 1.8$ kpc, evidenced from its direct measurement in edge-on galaxies \citep{BScaleHeightKrause}, in agreement with its value for the Milky Way, $\sim 1.5$ kpc \citep{BScaleHeightBeck}. Therefore, the magnetic field scale height can be as large as ~6--7 kpc. Since the LMC is a smaller galaxy (with radius $\sim 3$ times smaller than the Milky Way's), we use in our calculations the value $z_0=2$ kpc. 

Let us notice that the exponential decay in eq.~(\ref{BExponential}) is a conservative choice and can only underestimate the value of the field, since the value of the magnetic field obtained through the RM is already mediated along the line-of-sight.

\section{Dark Matter Density Profile} \label{profiles}
The intensity of the radio emission due to WIMPs annihilation coming from a given region within the LMC's disc depends on how these particles are distributed in the halo of the LMC. In fact, if $\rho(r)$ is the density of WIMPs as a function of galactocentric distance $r$, the emissivity at a give frequency will be proportional to $\rho^2(r)$. We need, therefore, to estimate $\rho(r)$ in order to estimate the emission due to DM annihilation.

In \citet{Angela}, different DM density profiles are studied using measurements of the LMC's rotational velocity field. Following this approach, we used the results obtained from HI \citep{H1RotationCurve} and carbon star data \citep{AlvesNelson} to estimate $\rho(r)$. The HI velocity field is composed of 26 data points at galactocentric distances between $\sim 0.05$ kpc and $\sim 3$ kpc. The carbon stars obtained from 422 stars results in 4 data points, with galactocentric distances between 4.0 kpc and 8.2 kpc. We fit to this data set six different DM density profiles suggested in the literature:

\begin{itemize}
\item A simple isothermal sphere cored profile: 
\begin{displaymath}
\rho_{{\rm iso}}(r)=\frac{\rho_0}{1+(\frac{r}{r_0})^2};
\end{displaymath}

\item the Navarro--Frenk--White (NFW) profile, derived from N-body simulations of CDM haloes for structures ranging from dwarf galaxies to clusters of galaxies \citep{NFW1,NFW2}, which diverges as $\rho\propto r^{-1}$ in the central region of the galaxy: 
\begin{displaymath}
\rho_{{\rm NFW}}(r)=\frac{\rho_0}{\frac{r}{r_0}\Big(1+\frac{r}{r_0}\Big)^2};
\end{displaymath}

\item the Moore et al. profile \citep{Moore}, derived from independent higher resolution CDM halo simulations, which has a steeper slope than the NFW profile for the central regions of the galaxy ($\rho\propto r^{-1.5}$): 
\begin{displaymath}
\rho_{{\rm M}}(r)=\frac{\rho_0}{\Big(\frac{r}{r_0}\Big)^{1.5}\Big(1+(\frac{r}{r_0})^{1.5}\Big)};
\end{displaymath}

\item the Burkert profile \citep{Burkert}, i.e. a cored profile derived from rotation curve fits of dwarf spiral galaxies: 
\begin{displaymath}
\rho_{{\rm B}}(r)=\frac{\rho_0}{\Big(1+\frac{r}{r_0}\Big)\Big(1+(\frac{r}{r_0})^2\Big)};
\end{displaymath}

\item the Hayashi et al. profile \citep{Hayashi}, which is a modified NFW profile derived from N-body simulations of the evolution of CDM subhaloes undergoing tidal stripping while orbiting around a larger halo: 
\begin{eqnarray}
\lefteqn{\rho_{{\rm H}}(r)=\frac{f_t}{1+(\frac{r}{r_0})^{3}}\,\,\rho_{{\rm NFW}}(r)  {} }
\nonumber\\
&& {}\,\,\,\,\,=\frac{\rho_0}{1+(\frac{r}{r_{0}})^{3}}\Bigg[ \frac{1}{\frac{r}{r_{0,{\rm NFW}}}\Big(1+\frac{r}{r_{0,{\rm NFW}}}\Big)^2} \Bigg],\label{Hayashi}
\nonumber
\end{eqnarray}
where $\rho_0$ $\equiv$ $f_t$. $\rho_{0,{\rm NFW}}$;

\item and, finally, the Einasto profile, with three free parameters, first introduced in 1965 by Einasto \citep{Einasto} to describe the distribution of stars in our galaxy, and later used to describe the results of N-body simulations of DM haloes \citep{NavarroEinasto}:
\begin{displaymath}
\rho_{{\rm E}}(r)=\rho_0\,{\rm exp}\bigg\{\frac{2}{\alpha}\bigg[\bigg(\frac{r}{r_0}\bigg)^{\alpha}-1\bigg]\bigg\}.
\end{displaymath}
 
\end{itemize}
The NFW, the Moore et al. and the Hayashi et al. profiles were already tested in \citet{Angela}.

The values of the free parameters were determined by fitting each profile to the rotation velocity field using that the velocity $V(r)$ at an orbit of radius $r$, inside of which a DM mass $M(r)$ is enclosed, is given by:
\begin{equation}
V(r)=\sqrt{\frac{GM(r)}{r}}. \label{rotation}
\end{equation}

\begin{figure}
\begin{center}
\includegraphics[width=8cm]{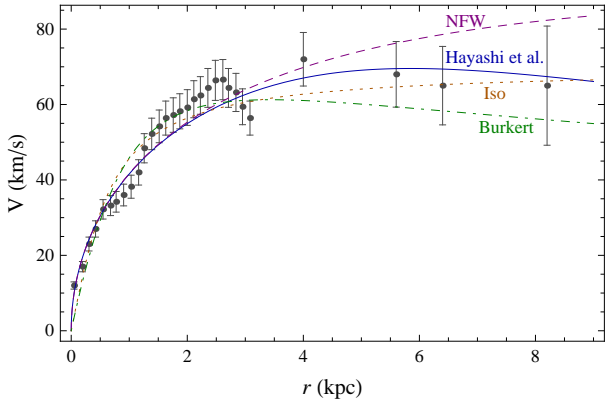}  
\caption{Velocity field data points superimposed with the resulting fitted curves for the isothermal sphere, NFW, Burkert and Hayashi et al. density profiles.}  \label{Fits}
\end{center}
\end{figure}

\begin{figure}
\begin{center}
\includegraphics[width=8cm]{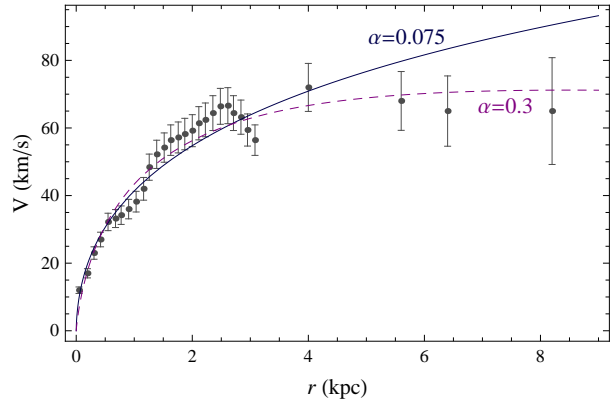}  
\caption{Results of fits for the Einasto density profile for $\alpha=0.075$, which resulted in the smallest value of $\chi^{2}/d.o.f.$, and $\alpha=0.3$.}  \label{FitEinasto}
\end{center}
\end{figure}

The Moore et al. profile gave the worst fit to the data, with a $\chi^{2}/d.o.f.\sim8$. The fitted curves for the other profiles can be seen in Figs.~\ref{Fits} and~\ref{FitEinasto} and the resulting values for the free parameters for each density profile are displayed in Table~\ref{FitResults}, as well as the value of the corresponding $\chi^{2}/d.o.f.$ For the purpose of fitting the Hayashi et al. profile, we fixed $r_{0,NFW}=9.04$ kpc, its best-fitting value previously found for the NFW profile.

\begin{table}
\caption{Values of the parameters $\rho_0$ and $r_0$ and $\chi^2/d.o.f.$ obtained from the fit to rotation curve data for each profile.}  \label{FitResults}
\begin{tabular}{@{}lccc@{}}
\hline
Profile & $\rho_0\,$($\times 10^6\, {\rm M}_\odot\,$kpc$^{-3}$) & $r_0\,$(kpc) & $\chi^2/d.o.f.$\\
\hline
NFW & 8.18 $\pm$ 2.67 & 9.04 $\pm$ 2.43 & 1.26\\
Iso. Sphere & 326 $\pm$ 65 & 0.52 $\pm$ 0.076 & 4.23\\
Burkert & 289 $\pm$ 51  & 1.06 $\pm$ 0.13 & 4.54\\
Hayashi et al. & 8.16 $\pm$ 0.30 &  6.36 $\pm$ 2.08 & 1.17\\
Einasto & 0.0015 $\pm$ 0.0008 & 824 $\pm$ 381 & 1.52\\
\hline
\end{tabular}
\end{table}

We can see that both the isothermal sphere and the Burkert profile did not fit well the data, resulting in $\chi^{2}/d.o.f.>4$.

For the Einasto profile, we found that good fits (with values of $\chi^{2}/d.o.f.$ around 1.5) can be obtained for values of $\alpha$ between $0.06$ and $0.1$. The smallest value of $\chi^{2}/d.o.f.$ was obtained for $\alpha=0.075$ and is displayed in Table~\ref{FitResults}. We can see in Fig.~\ref{FitEinasto}, that the resulting curve for $\alpha=0.075$ fits very well the rotation curve data for small values of the radius, but does not reproduce its behaviour at larger galactocentric distances. We also plot in Fig.~\ref{FitEinasto} the curve obtained with $\alpha=0.3$. In this case, the fit results in $\rho_{{\rm E}} = 6.38 \pm 1.32 \times 10^6$  M$_\odot$ kpc$^{-3}$ and $r_{{\rm E}}=4.16 \pm 0.67$ kpc. Although this curve reproduces better the global behaviour of the data, we get a larger value of $\chi^{2}/d.o.f.=2.05$. This happens because the small radius points (with smaller error bars) are better fitted by the curve with $\alpha=0.075$. 

\begin{figure}
\begin{center}
\includegraphics[width=8cm]{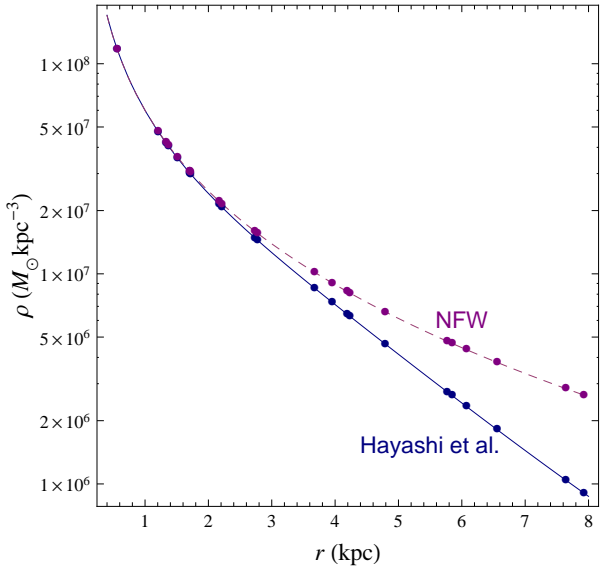}  
\caption{NFW and Hayashi et al. density profiles as a function of $r$ for the best-fitting parameters. The dots show the positions of all the 23 regions.}  \label{NFWHayashi}
\end{center}
\end{figure}

In any case, the NFW and the Hayashi et al. profiles fit the rotation curve better than the Einasto profile. The behaviour of these two density profiles (with the values of the free parameters set to the ones in Table~\ref{FitResults}) as a function of the galactocentric distance $r$ is shown in Fig.~\ref{NFWHayashi}. We can see that the DM densities predicted by the two profiles are practically the same for $r \la 2$ kpc. 

Nevertheless, at the largest distance sampled by our background sources, 7.92 kpc, the difference between the two density profiles is much smaller than one order of magnitude: $\rho_{{\rm NFW}}(7.92\,\,{\rm kpc})\sim2.8\,\rho_{{\rm H}}(7.92\,\,{\rm kpc})$.

We therefore chose to use the Hayashi et al. density profile in our calculations, both because its fit to the data resulted in the lowest value of $\chi^{2}/d.o.f.$ and because it provides a DM density value lower than the one provided by the NFW profile for most of the regions we are interested in (conservative choice).

\section{Energy loss processes} \label{EnergyLosses}
While diffusing away from the production site, several energy loss processes act on the $e^+e^-$ fluid. Our calculations accounted only for the three fastest energy loss processes: the synchrotron emission, the inverse Compton scattering (ICS) off the background photons and bremsstrahlung. We do not include Coulomb energy losses. 

\begin{figure*}
\begin{center}
\includegraphics[width=17cm]{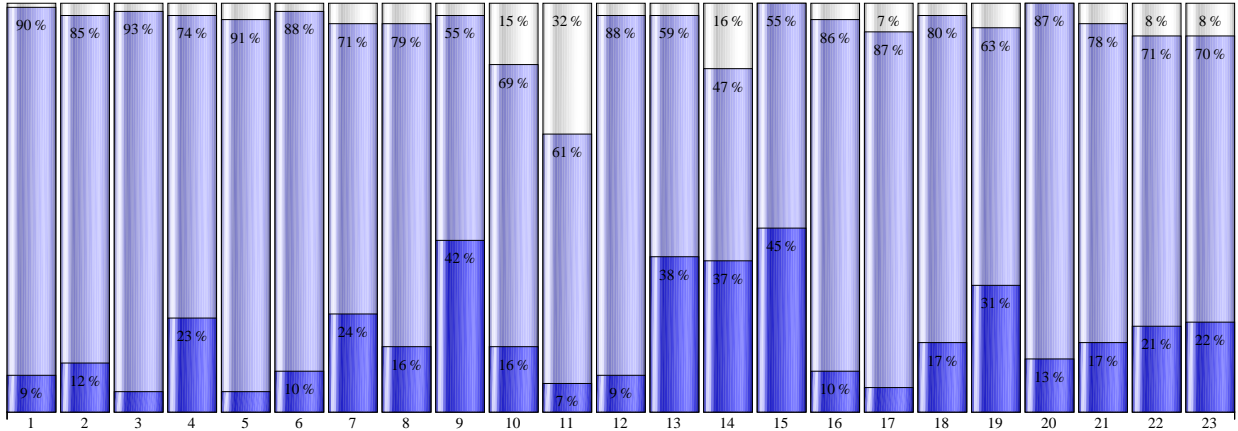}
\caption{Relative contribution of each energy loss process calculated at the position of each region at 1.4 GHz (see text). The darkest and lightest colours (bottom and upper parts) show the synchrotron and the bremsstrahlung contributions, respectively. The middle parts show the ICS contribution.} \label{EnergyLoss}
\end{center}
\end{figure*}

\subsection{Synchrotron}
An electron or a positron with energy $E_e$ diffusing in a region filled with magnetic field $B$ will lose energy via synchrotron radiation with the following characteristic time-scale:
\begin{displaymath}
\tau_{{\rm syn}}=\tau^{0}_{{\rm syn}} \bigg(\frac{B}{\mu {\rm G}} \bigg)^{-2}\, \bigg(\frac{E_e}{{\rm GeV}} \bigg)^{-1},
\end{displaymath}
where $\tau^{0}_{{\rm syn}}=3.95 \times 10^{17}$ s.

\subsection{Inverse Compton Scattering}
Similarly, the characteristic time for the ICS is given by
\begin{displaymath}
\tau_{{\rm ICS}}=\tau^{0}_{{\rm ICS}} \bigg(\frac{U_{{\rm rad}}}{{\rm eV cm}^{-3}} \bigg)^{-1}\, \bigg(\frac{E_e}{{\rm GeV}} \bigg)^{-1},
\end{displaymath}
where $\tau^{0}_{{\rm ICS}}=9.82 \times 10^{15}$ s and $U_{{\rm rad}}$ is the energy density of the Interstellar Radiation Field (ISRF). 

In order to estimate the intensity of the ISRF at the position of each region, we used the map of dust temperature ($T_d$) in the LMC presented in \citet{Bernard}. The energy density of the ISRF is proportional to $T_d^{4+\beta}$. We used $\beta=2$ \citep{Boulanger}. To obtain the proportionality constant, we use the fact that the dust temperature in the solar neighbourhood is 17.5 K and the local value of the ISRF is 0.539 eV cm$^{-3}$ \citep{WeingartnerDraine}. We can therefore estimate the the ISRF at the position of each one of the 23 regions we are interested in.

Following the same conservative approach we adopted in the case of the magnetic field, we impose that the value of the ISRF obtained in this way corresponds to its value on the disc, $U_{{\rm rad}}^{{\rm disc}}$, and assume that, for each region, $U_{\rm rad}$ decays exponentially with $z$:
\begin{equation}
U_{{\rm rad}}(z)=U_{{\rm rad}}^{{\rm disc}} \,{\rm e}^{-|z|/h_{0}}, \label{UradEquation}
\end{equation}
where $h_{0}$ is the disc scale height, which is actually a function that grows with $r$ and can be modelled as \citep{AlvesNelson}:
\begin{equation}
h_{0} (r) = h_{0} (0)\,\, {\rm e}^{r/\xi}\,\, {\rm kpc}, \label{ScaleHeight}
\end{equation}
where $h_{0}(0)=0.14$ kpc is the value of the disc scale height in the centre of the LMC and $\xi=2.24$ kpc.

Table~\ref{TableRCs} shows the values of $U_{{\rm rad}}^{{\rm disc}}$ for all the 23 regions under study.

\subsection{Bremsstrahlung}
The characteristic time for the bremsstrahlung process is given by \citep{Longair}:
\begin{displaymath}
\tau_{{\rm brem}}=\tau^{0}_{{\rm brem}} \Big(\frac{n_{{\rm H}}}{{\rm cm}^{-3}} \Big),
\end{displaymath}
where $\tau^{0}_{{\rm brem}}=1.17 \times 10^{15}$ s and $n_{\rm H}$ is the hydrogen number density. 

To estimate the value of $n_{\rm H}$ in our selected regions, we used the table of HI clouds in the LMC presented in \citet{CloudTable}, which contains the coordinates, radius and mass of each cloud. We assumed that the clouds are spherical and estimated their density by simply dividing their mass by the volume of a sphere with the reported radius. We then checked if each one of the 23 regions fell inside a cloud and, if so, associated to that region the density of the cloud. If a region fell inside more than one cloud, we associated to that region the density of the highest density cloud, which is a conservative choice. Regions that did not fall inside any cloud were assigned with $n_{\rm H}=0$. The values of $n_{\rm H}$ for each region obtained in this way are reported in Table~\ref{TableRCs}.

We neglect the contribution of ionised hydrogen in our calculations. This is a reasonable approximation in view of the depolarisation results obtained in \citet{MagField}, which indicate that there are no bright individual HII regions in the directions where the RMs were observed.\newline

To include the bremsstrahlung in the calculations, we need to modify the $\mu({\bf r})$ function defined in \citet{PaperMW} to:
\begin{eqnarray}
\lefteqn{\mu\,({\bf r})= \Bigg[\,\frac{\tau^0_{\rm syn}}{\tau^0_{\rm ICS}} \frac{U_{\rm rad}({\bf r})}{{\rm eV\,cm}^{-3}} + \frac{\tau^0_{\rm syn}}{\tau^0_{\rm brem}} \sqrt{\frac{\nu_0}{\nu}\frac{B({\bf r})}{\mu{\rm G}}} \frac{n_{\rm H}}{{\rm cm}^{-3}}+ {} }
\nonumber\\
&& {}\,\,\,\,\, +\Big(\frac{B({\bf r})}{\mu{\rm G}}\Big)^2 \Bigg]^{-1},
\nonumber
\end{eqnarray}
where $\nu$ is the frequency and $\nu_0=3.7 \times 10^6$ Hz.

Fig.~\ref{EnergyLoss} compares the energy loss rate, $b=E_e/\tau$, for each of the three energy loss processes, at the position of all the 23 regions. For each region, we plot the values of the normalised energy loss rate for the synchrotron (calculated using $B_{\rm reg}$), ICS and the bremsstrahlung processes. All the values were evaluated on the plane of the disc ($z=0$) and we used the frequency peak approximation $E_e=\sqrt{B_{\rm reg}\times\nu/\nu_0}$ GeV, where $B_{\rm reg}$ is measured in $\mu$G. Since the results were very similar for $\nu=1.4$ and $\nu=4.8$ GHz, we show only the results obtained for 1.4 GHz.

While ICS is clearly the dominant process in all regions we are studying, the contribution of bremsstrahlung seems to be the the smallest one. It is however, surely not negligible in at least 3 regions (10, 11 and 14). These processes therefore, cannot be neglected. 

If we remember that the regions are numbered according to their galactocentric distance, we immediately notice in Fig.~\ref{EnergyLoss} the apparent lack of correlation between the energy loss rates (and therefore of the magnetic field, the ISRF and the hydrogen distribution) and the galactocentric distance, which is clearly a consequence of the irregular nature of the LMC.

\section{Results} \label{ImposingConstraints}
We considered two possible WIMPs annihilation channels: $\chi \chi \to b \overline{b}$, in which electrons and positrons will be produced by decaying muons ($\mu^- \to e^-\, \overline{\nu}_{e} \nu_{\mu}$) and anti-muons ($\mu^+ \to e^+\, \overline{\nu}_{\mu} \nu_{e}$) produced in pions decays ($\pi^- \to \mu^- \overline{\nu}_{\mu}$ and $\pi^+ \to \mu^+ \nu_{\mu}$) and the leptophilic channel $\chi \chi \to \mu^+ \mu^-$.

\begin{figure*}
\subfigure{
\includegraphics[scale=1.0]{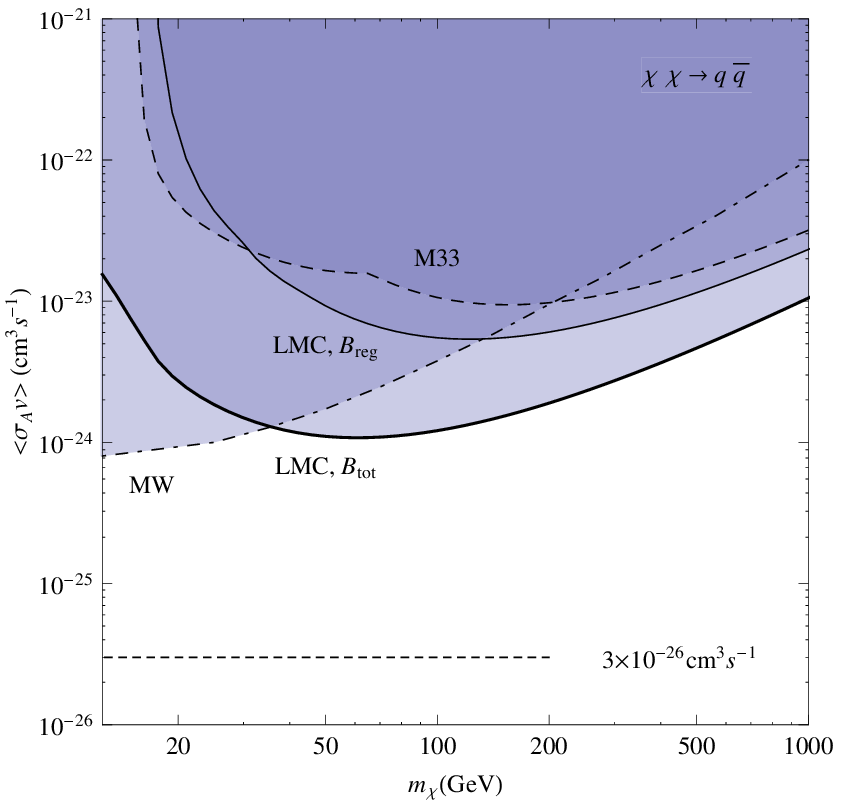}\label{Quarks}}
\subfigure{
\includegraphics[scale=1.0]{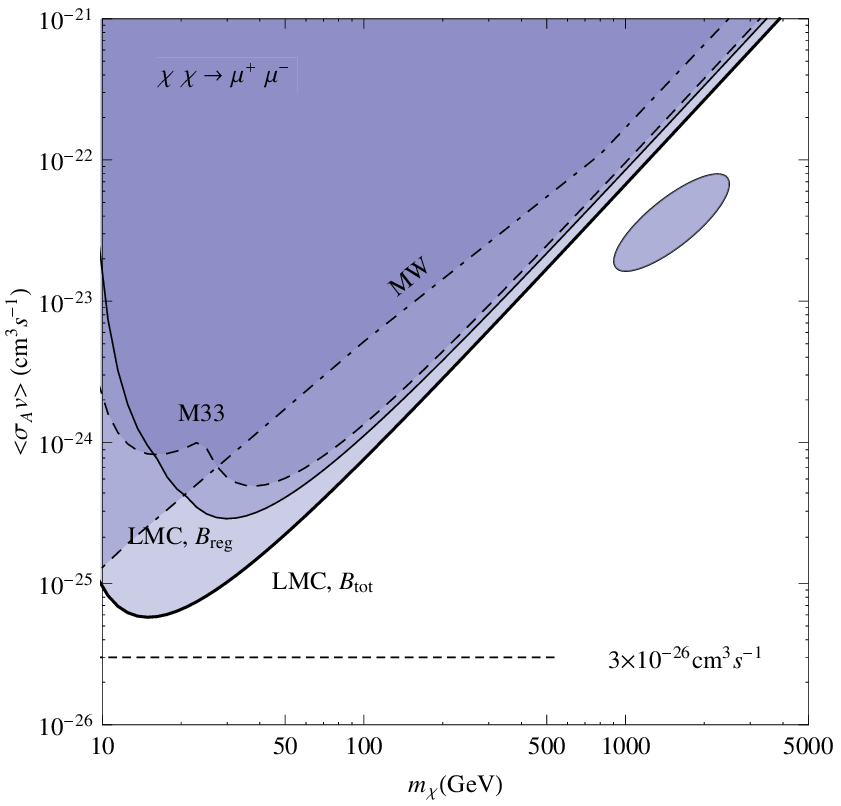}\label{Leptons}}
\caption{Constraints obtained with the $\chi \chi \to b\overline{b}$ channel (left) and the $\chi\chi \to \mu^+ \mu^-$ channel (right). The solid lines indicate the best constraints obtained with the LMC using either $B_{\rm tot}$ or $B_{\rm reg}$. The dashed lines indicate the best constraints obtained with M33 \protect\citep{PaperM33} and the dot-dashed ones indicate those obtained with the Milky Way \protect\citep{PaperMW}. The shaded regions are the forbidden ones. The ellipse shaped region indicates the regions favoured by the Pamela/HESS/Fermi--LAT results \protect\citep{Meade}.} \label{Results1}
\end{figure*}

Leptophilic channels have recently raised interest in view of the experimental results on the electron/positron cosmic ray spectra. While Pamela observed an unexpected rise in the positron fraction \citep{PAMELA}, Fermi--LAT observes a deviation from a simple power-law spectrum \citep{b1}, thus confirming the previous results obtained by HESS \citep{Hess}. If these results are to be interpreted as due to DM annihilation in the galactic halo, one needs to consider leptophilic channels and high mass scales. 

\begin{figure*}
\begin{center}
\subfigure{
\includegraphics[scale=1.0]{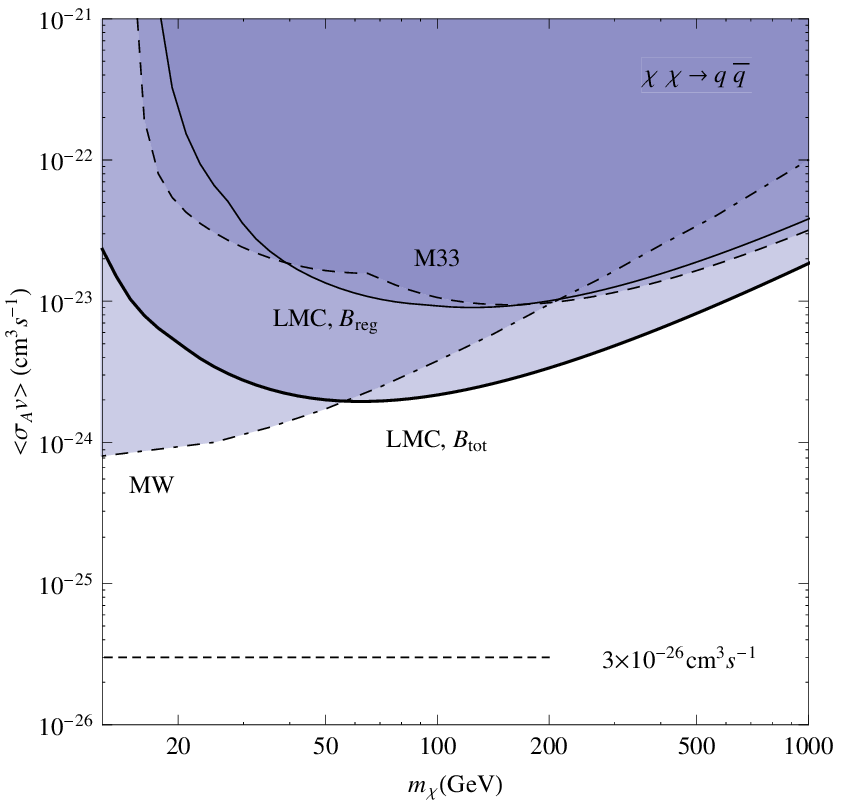}\label{QuarksISO}}
\subfigure{
\includegraphics[scale=1.0]{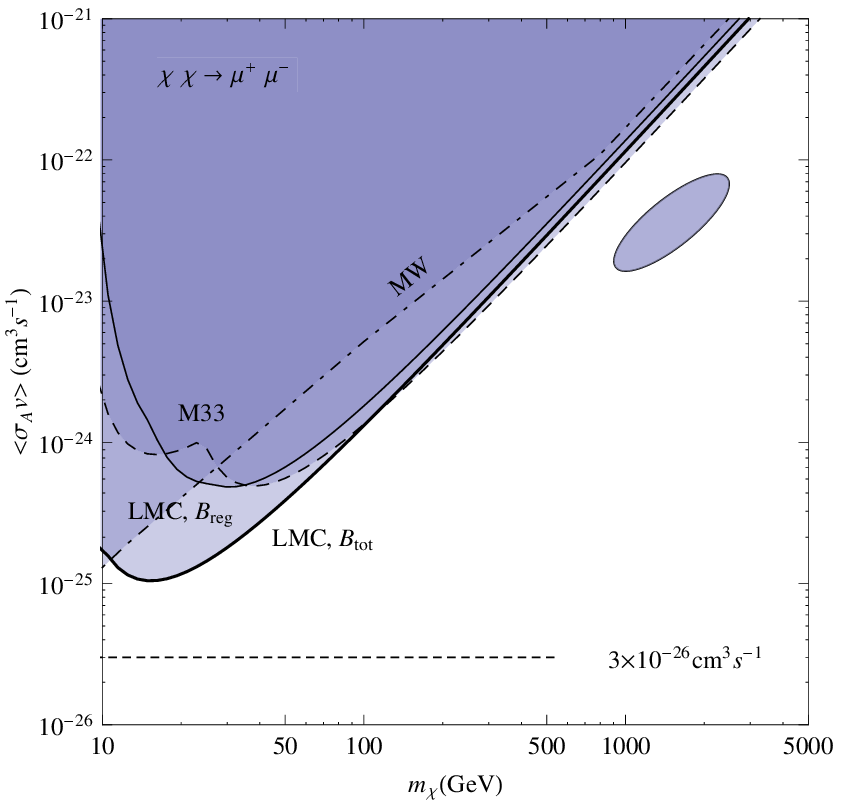}\label{LeptonsISO}}
\caption{Same as Fig.~\ref{Results1} using the isothermal sphere DM density profile with the parameters obtained in \protect\citet{AlvesNelson}.} \label{Results2}
\end{center}
\end{figure*}

The synchrotron intensity at a frequency $\nu$ due to DM annihilation coming from a region inside a solid angle $d\Omega$ on the LMC's disc is:
\begin{equation}
\frac{dI}{d\Omega} = \frac{{\rm cos}\,i}{4\pi} \int j_{\nu}(r,z) ds, \label{SynFlux}
\end{equation}
where $i$ is the disc inclination, $s$ is the distance along the line-of-sight and $j_{\nu}(r,z)$ is the synchrotron emissivity at a position in the LMC's halo with galactocentric distance $r$ along the disc and height $z$ above or below the disc. The term cos$\,i$ accounts for the fact that the line-of-sight is not parallel to $z$.

The expression of the emissivity is derived in detail in \citet{PaperMW}. 

We evaluated eq. (\ref{SynFlux}) for all 23 regions using $\nu=1.4$ and 4.8 GHz. The most constraining results were obtained for 1.4 GHz and can be seen in Fig.~\ref{Results1} for both annihilation channels considered. For comparison, we show also the best constraint obtained in \citet{PaperM33} for M33 using an NFW profile and assuming equipartition between magnetic fields and cosmic rays, and the best constraint obtained in \citet{PaperMW} for the Milky Way using the $\chi \chi \to b \overline{b}$ channel. For the $\chi \chi \to \mu^+ \mu^-$ channel we show the best constraint obtained for the Milky Way using the same formalism and observations described in \citet{PaperMW} and also the favoured region obtained when one attributes to DM annihilation the experimental results described above \citep{Meade}.

\section{Summary and Discussion} \label{Discussion}
We have imposed constraints on the $m_{\chi}$-$\langle\sigma_Av\rangle$ plane using radio observations at 1.4 GHz and 4.8 GHz of the LMC and analysing two different DM annihilation channels, a hadronic and a leptonic one. 

The existence of high resolution observations of the LMC in several frequency bands has allowed us to obtain most of the information needed to calculate the DM annihilation signal, making the least possible number of hypotheses in all the steps of the calculation. Being able to escape from this problem and, when necessary, making conservative assumptions, we have, therefore, obtained very robust results. 

In all cases studied, the best constraints were obtained using 1.4 GHz, since the number of $e^-$ and $e^+$ produced by DM annihilation decreases with energy. For higher frequencies to produce competing constraints, the observed intensity at those frequencies should be much lower than at 1.4 GHz. 

From Fig.~\ref{Results1}, we can see that the constraints on the $m_{\chi}$-$\langle\sigma_Av\rangle$ plane imposed from the analysis of the LMC are stronger than the ones obtained with M33 and with the Milky Way over the most of mass range considered when, excluding very low mass regions. 

In \citet{Angela} the estimated DM annihilation signal at different frequencies was obtained, fixing $m_{\chi}=50$ GeV and $\langle\sigma_Av\rangle=2 \times 10^{-26}$ cm$^3\,$s$^{-1}$ and using a constant magnetic field of intensity either $B=5\, \mu\,$G or $B=18.4\, \mu\,$G. We can estimate from their fig. 5 that at 1.4 GHz they obtain $\langle\sigma_Av\rangle \la 4 \times 10^{-25}$ cm$^3\,$s$^{-1}$ for $B=5\, \mu\,$G and $\langle\sigma_Av\rangle \la 3 \times 10^{-25}$ cm$^3\,$s$^{-1}$ for $B=18.4\, \mu\,$G. From our Fig.~\ref{Results1}, we can see that for $m_{\chi}=50$ GeV our best constraints using the hadronic channel are $\langle\sigma_Av\rangle <\, 9.25 \times 10^{-24}$ cm$^3\,$s$^{-1}$ using only $B_{\rm reg}$ and $\langle\sigma_Av\rangle <\, 1.10 \times 10^{-24}$ cm$^3\,$s$^{-1}$ using $B_{\rm tot}$. This result can be explained by the hypotheses adopted in \citet{Angela}, which have overestimated the DM annihilation signal. In the first place, inspection of Table~\ref{TableRCs}, shows that imposing a constant magnetic field of $18.4\, \mu\,$G for the entire volume of the LMC is not realistic. Even if we take into account only the $B_{\rm tot}$ column, we see that more than half of the regions present $B_{\rm tot} < 10\, \mu \,$G. As for the hypothesis $B=5\, \mu\,$G, we can see from Fig.~\ref{EnergyLoss} that the bremsstrahlung and, in particular, the ICS processes cannot be neglected (see for example regions 9 and 15, for which $B_{\rm reg}=5.32\, \mu\,$G and $4.57\, \mu \,$G, respectively). In fact, if we do neglect these two processes in our calculations, our results improve by a factor $\sim 3$.

Finally, we would like to comment on one possible shortcoming of our analysis. When fitting the LMC's rotation curve, we took into consideration only the DM mass, ignoring the luminous matter. To understand the effect of this choice, we redid our calculations using the results obtained in \citet{AlvesNelson}, where an isothermal sphere density profile is fitted to the LMC's rotation curve after subtracting the disc component (stars + gas). The resulting values of the free parameters are $\rho_{iso}= 10^8 {\rm M}_{\odot}$ kpc$^{-3}$ and $r_{iso}=1$ kpc. Using this density profile, the best constraints were obtained at 1.4 GHz and can be seen in Figs.~\ref{QuarksISO} and~\ref{LeptonsISO}.  

We can see that, accounting for the luminous matter in this way, our results are not very much affected and the conclusions previously described are still valid.

\section*{Acknowledgments}
We would like to thank Maurizio Paolillo for fruitful discussions and Jean-Philippe Bernard, for kindly sending us the dust temperature data. G.M. would like to thank the partial support of MIUR - PRIN2008 `Fisica Teorica Astroparticellare'.

\label{lastpage}

\end{document}